\documentclass[draft,showpacs,showkeys,eqsecnum,nofootinbib,aps]{revtex4}
\renewcommand{\theequation}{\arabic{equation}}
\def\beq{\begin{equation}}
\def\eeq{\end{equation}}
\def\bea{\begin{eqnarray}}
\def\eea{\end{eqnarray}}
\def\nn{\nonumber}

\def\pa{\partial}

\begin{document}
\title{BRST symmetries in free particle system on toric geometry}
\author{Soon-Tae Hong}
\email{soonhong@ewha.ac.kr} \affiliation{Department of Science
Education, Ewha Womans University, Seoul 120-750 Korea}
\date{\today}%
\begin{abstract}
We study a free particle system residing on a torus to investigate its
Becci-Rouet-Stora-Tyutin symmetries associated with its St\"uckelberg
coordinates, ghosts and anti-ghosts.  By exploiting zeibein frame on the
toric geometry, we evaluate energy spectrum of the system to describe
the particle dynamics.  We also investigate symplectic structures involved
in the second-class system on the torus.
\end{abstract}
\maketitle

\section{Introduction}
\setcounter{equation}{0}
\renewcommand{\theequation}{\arabic{section}.\arabic{equation}}

It is well known in string theory that toric geometry is a generalization
of the projective identification that defines $CP^{n}$ corresponding to
the most general linear sigma model, and it provides a scheme for
constructing Calabi-Yau manifolds and their mirrors~\cite{pol}.  Recently,
on the basis of boundary string field theory~\cite{witten92}, the brane-antibrane
system was exploited~\cite{hotta} in the toroidal background to investigate
its thermodynamic properties associated with the Hagedorn temperature~\cite{witten88,witten01}.
The Nahm transform and moduli spaces of $CP^{n}$ models were also studied
on the toric geometry~\cite{aguado02}.  In a four dimensional, toroidally
compactified heterotic string, the electrically charged BPS-saturated states were
shown to become massless along the hypersurfaces of enhanced gauge symmetry of a
two-torus moduli subspace~\cite{cvetic96}.

The Becci-Rouet-Stora-Tyutin (BRST) symmetries~\cite{brst} have been constructed for
constrained systems~\cite{hongpr} in the Batalin-Fradkin-Vilkovisky (BFV) scheme~\cite{bfv}.
In order to treat rigorously the constraints involved in the systems, the
Dirac quantization was proposed~\cite{di}, and it was later improved by converting a
second-class constraint system into a first-class one in the Batalin-Fradkin-Tyutin
(BFT) embedding~\cite{bft}, where the BRST symmetries can be generated.

Recently, the gauge symmetry enhancement was studied~\cite{hong03prd}
on target space of Grassmann manifold in the Dirac Hamiltonian formalism, and the
BRST invariant effective Lagrangian was also realized in noncommutative D-brane
system with NS $B$-field~\cite{hongbrst}.  To show novel phenomenological
aspects~\cite{cs}, the compact form of the first-class Hamiltonian has been also
constructed~\cite{hong99o3} for the O(3) nonlinear sigma model, which has been
also studied~\cite{rothe03} to investigate the symplectic structures~\cite{faddeev88}
and BFT embeddings.

In this paper, we will construct a first-class Hamiltonian by
introducing the St\"uckelberg coordinates associated with the
geometrical constraints on the torus.  In the BFV scheme, we will then find a
BRST-invariant gauge fixed Lagrangian including ghosts and anti-ghosts, and
the corresponding BRST transformation rules.  We will
also construct the spectrum and the symplectic structures of the free particle system
on the torus.  In Sec. II, we will introduce a free particle system residing on a
torus, to construct the first-class Hamiltonian and Dirac brackets.  In Sec. III,
we will introduce canonical sets of ghosts and anti-ghosts in the BFV scheme,
to construct the BRST symmetric effective Lagrangian.  In Sec. IV, we will obtain the spectrum
of the free particle on the torus to figure out the particle dynamics.  In Sec. V, we will
study the symplectic structures involved in the second-class system on the torus.

\section{First-class constraints and Hamiltonian}
\setcounter{equation}{0}

\renewcommand{\theequation}{\arabic{section}.\arabic{equation}}

In this section, we consider a free particle system residing on a torus,
whose Lagrangian is of the form
\begin{equation}
L_{0}=\frac{1}{2}m\dot{r}^{2}+\frac{1}{2}mr^{2}\dot{\theta}^{2}
+\frac{1}{2}m(b+r\sin\theta)^{2}\dot{\phi}^{2}.
\label{lag}
\end{equation}
where we have used toroidal coordinates $(r,\theta,\phi)$ for toric geometry
\beq
x_{1}=(b+r\sin\theta)\cos\phi,~~~
x_{2}=(b+r\sin\theta)\sin\phi,~~~
x_{3}=r\cos\theta,
\label{toco1}
\eeq
to satisfy
\beq
((x_{1}^{2}+x_{2}^{2})^{1/2}-b)^{2}+x_{3}^{2}=r^{2}.
\label{torusgeo}
\eeq
Note that we have now a torus with axial circle in the $x_{1}$-$x_{2}$ plane centered at the
origin, of radius $b$, having a circular cross section of radius $r$,
and the angle $\theta$ ranges from $0$ to $2\pi$, and the angel $\phi$ from $0$ to $2\pi$.
To fulfill the toric geometry (\ref{torusgeo}), we can
also exploit anther toroidal coordinates $(\mu,\eta,\phi)$ defined as~\cite{mo}
\beq
x_{1}=\frac{c\sinh\mu\cos\phi}{\cosh\mu-\cos\eta},~~~
x_{2}=\frac{c\sinh\mu\sin\phi}{\cosh\mu-\cos\eta},~~~
x_{3}=\frac{c\sin\eta}{\cosh\mu-\cos\eta},
\label{toco2}
\eeq
where $\mu$ ranges from $0$ to $\infty$, $\eta$ from $0$ to $2\pi$,
and $\phi$ from $0$ to $2\pi$.  Here, we have relations between these two
coordinate systems (\ref{toco1}) and (\ref{toco2}),
\beq
r=\frac{c}{\sinh\mu},~~~
\theta=\cos^{-1}\frac{\sinh\mu\sin\eta}{\cosh\mu-\cos\eta},~~~
\phi=\phi,~~~
b=c{\rm coth}\mu.
\eeq

Now, we impose the condition that the particle is constrained
to satisfy a geometrical constraint
\begin{equation}
\Omega_{1}=r-a\approx 0.
\label{c1}
\end{equation}
By performing the Legendre transformation, one can obtain the canonical
Hamiltonian,\footnote{One can include the constraint (\ref{c1}) explicitly in the Lagrangian
to yield $L=L_{0}+u(r-a)$ with a Lagrangian multiplier $u$.  One can then obtain
a primary constraint $\Omega_{0}=p_{u}$ with $p_{u}$ being momentum conjugate
to $u$.  The Hamiltonian is then given by $H_{T}=H_{0}-u(r-a)$ and successive time evolutions of
$\Omega_{0}$ reproduce $\Omega_{1}=r-a$ and $\Omega_{2}=p_{r}$.  The condition
$\{\Omega_{2},H_{T}\}=0$ fixes value of $u$, namely $u=-p_{\theta}^{2}/(mr^{3})
-p_{\phi}^{2}\sin\theta/(m(b+r\sin\theta)^{3})$, which can terminate series of
constraints. Since $\Omega_{0}$ is first-class, one can thus end up with two second-class
constraints $\Omega_{1}$ and $\Omega_{2}$, which are used in the context.}
\begin{equation}
H_{0}=\frac{p_{r}^{2}}{2m}+\frac{p_{\theta}^{2}}{2mr^{2}}+\frac{p_{\phi}^{2}}{2m(b+r\sin\theta)^{2}}
\label{hc}
\end{equation}
where $p_{r}$, $p_{\theta}$ and $p_{\phi}$ are the canonical momenta conjugate
to the coordinates $r$, $\theta$ and $\phi$, respectively,
given by
\beq
p_{r}=m\dot{r},~~~p_{\theta}=mr^{2}\dot{\theta},~~~p_{\phi}=m(b+r\sin\theta)^{2}\dot{\phi}.
\label{cojm}
\eeq
The time evolution of the constraint $\Omega_1$ yields an additional secondary
constraint
\begin{equation}
\Omega_{2}=p_{r}\approx 0
\label{const22}
\end{equation}
and $\Omega_{1}$ and $\Omega_{2}$ form a second-class constraint algebra
\begin{equation}
\Delta_{kk^{\prime}}=\{\Omega_{k},\Omega_{k^{\prime}}\}
=\epsilon_{kk^{\prime}}\label{delta}
\end{equation}
with $\epsilon_{12}=-\epsilon_{21}=1$.  Since the constraints are second-class, we
can define Dirac bracket
\begin{equation}
\{F,G\}_{D}=\{F,G\}-\{F,\Omega_{k}\}\Delta^{k k^{\prime}}
\{\Omega_{k^{\prime}},G\}
\end{equation}
with $\Delta^{k k^{\prime}}$ being the inverse of $\Delta_{k k^{\prime}}$ in (\ref{delta})
to yield
\beq
\{r,p_{r}\}_{D}=0,~~~\{\theta,p_{\theta}\}_{D}=1,~~~\{\phi,p_{\phi}\}_{D}=1.
\label{dbracket}
\eeq
In the forthcoming section, the Dirac brackets (\ref{dbracket}) will be discussed in terms of
symplectic brackets.  In the quantum level, these Dirac brackets produce the
following commutator relations
\beq
[r,p_{r}]=0,~~~[\theta,p_{\theta}]=i\hbar,~~~[\phi,p_{\phi}]=i\hbar.
\label{qcomm}
\eeq

Following the BFT embedding~\cite{bft} which systematically
converts the second-class constraints into first-class ones, we introduce the
St\"uckelberg coordinates $(\eta,p_{\eta})$ with the Poisson brackets
$$
\{\eta, p_{\eta}\}=1,
$$
to obtain the first-class constraints as follows
\begin{eqnarray}
\tilde{\Omega}_{1}&=&\Omega_{1}-\eta=r-a-\eta,
\nonumber \\
\tilde{\Omega}_{2}&=&\Omega_{2}+p_{\eta}
=p_{r}+p_{\eta}.
\label{1stconst}
\end{eqnarray}
Note that these first-class constraints yield a strongly involutive first-class constraint algebra
$\{\tilde{\Omega}_{i},\tilde{\Omega}_{j}\}=0$, which is related with
the first Dirac bracket in (\ref{dbracket}) and that the particle is geometrically
constrained to reside on the torus with the modified radius $r=a+\eta$ in the extended phase space.

Next, we construct the first-class Hamiltonian $\tilde{H}_{0}$ as a power series in the
St\"uckelberg coordinates $(\eta,p_{\eta})$ by demanding that they are strongly involutive:
$\{\tilde{\Omega}_{i}, \tilde{H}_{0}\}=0$.  After some algebra, we obtain the first-class Hamiltonian,
\begin{equation}
\tilde{H}_{0}=\frac{(p_{r}+p_{\eta})^{2}}{2m}+\frac{p_{\theta}^{2}}{2m(r-\eta)^{2}}
+\frac{p_{\phi}^{2}}{2m(b+(r-\eta)\sin\theta)^{2}}.
\label{htilde}
\end{equation}
A problem with $\tilde{H}_{0}$ in (\ref{htilde}) is that it does not naturally
generate the first-class Gauss law constraint from the time evolution of the
constraint $\tilde{\Omega}_{1}$.  By introducing an additional term
proportional to the first-class constraints
$\tilde{\Omega}_{2}$ into $\tilde{H}_{0}$, we obtain an equivalent first-class
Hamiltonian
\begin{equation}
\tilde{H}=\tilde{H}_{0}-p_{\eta}\tilde{\Omega}_{2},
\label{hctp}
\end{equation}
which naturally generates the Gauss law constraint
\beq
\{\tilde{\Omega}_{1},\tilde{H}\}=\tilde{\Omega}_{2},~\{\tilde{\Omega}_{2},\tilde{H}\}=0.
\eeq

One notes here that $\tilde{H}_{0}$ and $\tilde{H}$ act in the same way
on physical states, which are annihilated by the first-class constraints.
Similarly, the equations of motion for observables remain unaffected by the
additional term in $\tilde{H}$.  Furthermore, in the limit
$(\eta,p_{\eta})\rightarrow 0$, our first-class system is exactly reduced to
the original second-class one.

\section{BRST symmetries and effective Lagrangian}
\setcounter{equation}{0}
\renewcommand{\theequation}{\arabic{section}.\arabic{equation}}

In this section, we introduce canonical sets of the ghosts and anti-ghosts
together with the Lagrangian multipliers in the BFV scheme~\cite{bfv}, which is
applicable to theories with the first-class constraints,
\[
({\cal C}^{i},\bar{{\cal P}}_{i}),~~({\cal P}^{i}, \bar{{\cal C}}_{i}),
~~(N^{i},B_{i}),~~~~(i=1,2),
\]
which satisfy the super-Poisson algebra
\[
\{{\cal C}^{i},\bar{{\cal P}}_{j}\}=\{{\cal P}^{i}, \bar{{\cal C}}%
_{j}\}=\{N^{i},B_{j}\}=\delta_{j}^{i}.
\]
Here the super-Poisson bracket is defined as
\[
\{A,B\}=\frac{\delta A}{\delta q}|_{r}\frac{\delta B}{\delta p}|_{l}
-(-1)^{\eta_{A}\eta_{B}}\frac{\delta B}{\delta q}|_{r}\frac{\delta A} {%
\delta p}|_{l},
\]
where $\eta_{A}$ denotes the number of fermions, called the ghost number,
in $A$ and the subscript $r$ and $l$ denote right and left derivatives,
respectively.

In our model, the nilpotent BRST charge $Q$ defined as
\beq
Q={\cal C}^{i}\tilde{\Omega}_{i}+{\cal P}^{i}B_{i},
\eeq
and the BRST invariant minimal Hamiltonian $H_{m}$ given by
\beq
H_{m}=\tilde{H}-{\cal C}^{1}\bar{{\cal P}}_{2},
\eeq
satisfy the relations
\begin{equation}
\{Q,H_{m}\}=0,~~Q^{2}=\{Q,Q\}=0.
\end{equation}
Our next task is to fix the gauge, which is crucial to identify the BFT
auxiliary coordinate $\eta$ with the St\"uckelberg coordinate.  The desired
identification follows if we choose the fermionic gauge fixing function
$\Psi$ as
\begin{equation}
\Psi=\bar{{\cal C}}_{i}\chi^{i}+\bar{{\cal P}}%
_{i}N^{i},
\end{equation}
with the unitary gauge
\begin{equation}
\chi^{1}=\Omega_{1},~~~\chi^{2}=\Omega_{2}.
\end{equation}
Here note that the $\Psi$ satisfies the following identity
\begin{equation}
\{\{\Psi,Q\},Q\}=0.
\end{equation}
The effective quantum Lagrangian is then described as
\begin{equation}
L_{eff}=p_{r}\dot{r}+p_{\theta}\dot{\theta}+p_{\phi}\dot{\phi}+p_{\eta}\dot{\eta}
+B_{2}\dot{N}^{2}+\bar{{\cal P}}_{i}\dot{{\cal C}}^{i}+\bar{{\cal C}}_{2} \dot{%
{\cal P}}^{2}-H_{tot}
\end{equation}
where $H_{tot}=H_{m}-\{Q,\Psi\}$ and the terms $B_{1}\dot{N}^{1}+\bar{{\cal C}}_{1}
\dot{{\cal P}}^{1}=\{Q,\bar{{\cal C}}_{1} \dot{N}^{1}\}$
have been suppressed by replacing $\chi^{1}$ with $\chi^{1} +\dot{N}^{1}$.

Now we perform path integration over the ghosts, anti-ghosts and Lagrangian
multipliers $\bar{{\cal C}}_{1}$, ${\cal P}^{1}$, $\bar{{\cal P}}_{1}$,
${\cal C}^{1}$, $B_{1}$ and $N^{1}$, by using the equations of motion.  This leads
to the effective Lagrangian of the form
\begin{eqnarray}
L_{eff}&=&p_{r}\dot{r}+p_{\theta}\dot{\theta}+p_{\phi}\dot{\phi}+p_{\eta}\dot{\eta}
+B\dot{N}+\bar{{\cal P}}\dot{{\cal C}}+\bar{{\cal C}}\dot{{\cal P}}
-\frac{(p_{r}+p_{\eta})^{2}}{2m}-\frac{p_{\theta}^{2}}{2m(r-\eta)^{2}}
-\frac{p_{\phi}^{2}}{2m(b+(r-\eta)\sin\theta)^{2}}
\nonumber \\
& &+(p_{r}+p_{\eta})N+Bp_{r}+\bar{{\cal P}}{\cal P}
\end{eqnarray}
with the redefinitions: $N=N^{2}$, $B=B_{2}$, $\bar{{\cal C}}=\bar{{\cal C}}_{2}$,
${\cal C}={\cal C}^{2}$, $\bar{{\cal P}}=\bar{{\cal P}}_{2}$, ${\cal P}={\cal P}_{2}$.

After performing the routine variation procedure and identifying
$N=-B-\dot{\eta}$ we arrive at the effective
Lagrangian of the form
\beq
L_{eff}=L_{0}+L_{WZ}+L_{gh},
\eeq
where $L_{0}$ is given by (\ref{lag}) and $L_{WZ}$ and $L_{gh}$ are given by
\begin{eqnarray}
L_{WZ}&=&\frac{1}{2}m\dot{\eta}(\dot{\eta}-2\dot{r})+\frac{1}{2}m\eta(\eta-2r)\dot{\theta}^{2}
+\frac{1}{2}m\eta\sin\theta(\eta\sin\theta-2b-2r\sin\theta)\dot{\phi}^{2}\nn\\
L_{gh}&=&\dot{B}\dot{\eta}+\dot{\bar{\cal C}}\dot{\cal C}.
\label{leffbrst}
\end{eqnarray}
The effective Lagrangian $L_{eff}$ is now invariant under the BRST-transformation
\begin{eqnarray}
\delta_{B}r&=&\lambda {\cal C},~~~~~~
\delta_{B}\theta=0,~~~\delta_{B}\phi=0,~~~\delta_{B}\eta=\lambda{\cal C},
\nonumber \\
\delta_{B}\bar{{\cal C}}&=&-\lambda B,~~~ \delta_{B}{\cal C}=0,~~~\delta_{B}B=0.
\end{eqnarray}

\section{Energy spectrum and particle dynamics}
\setcounter{equation}{0}
\renewcommand{\theequation}{\arabic{section}.\arabic{equation}}

Now, in order to investigate the energy spectrum of the free particle system
on a torus, we impose the first-class constraints (\ref{1stconst}) on the first-class
Hamiltonian $\tilde{H}_{0}$ in (\ref{htilde}) to yield
\beq
\tilde{H}_{0}=\frac{p_{\theta}^{2}}{2ma^{2}}+\frac{p_{\phi}^{2}}{2m(b+a\sin\theta)^{2}}.
\eeq
Since the free particle of interest is constrained to reside on the torus, we should include
the geometrical effects of the target manifold of the torus whose spatial two-metric is given by
\beq
ds^{2}=a^{2}d\theta^{2}+(b+a\sin\theta)^{2}d\phi^{2}.
\eeq
The natural choice of zweibein frame is then
\beq
e_{\theta}=\frac{1}{a}\frac{\pa}{\pa \theta},~~~
e_{\phi}=\frac{1}{b+a\sin\theta}\frac{\pa}{\pa \phi},
\eeq
to, together with the commutator relations (\ref{qcomm}), yield the Hamiltonian operator
\beq
\tilde{H}_{0}=-\frac{\hbar^{2}}{2m}\left[\frac{1}{a^{2}(b+a\sin\theta)}\frac{\pa}{\pa\theta}
\left((b+a\sin\theta)\frac{\pa}{\pa\theta}\right)
+\frac{1}{(b+a\sin\theta)^{2}}\frac{\pa^{2}}{\pa\phi^{2}}\right].
\label{toham}
\eeq
Note that in the $b\rightarrow 0$ limit, the Hamiltonian operator (\ref{toham}) on the torus reduces
to that on a two-sphere.\footnote{The two-sphere Laplacian is given by
$\frac{1}{r^{2}\sin\theta}\frac{\pa}{\pa\theta}(\sin\theta\frac{\pa}{\pa\theta})
+\frac{1}{r^{2}\sin^{2}\theta}\frac{\pa^{2}}{\pa\phi^{2}}$ in spherical coordinates
and it can be rewritten as $\pa_{i}\pa_{i}-\frac{2x_{i}}{x_{k}x_{k}}\pa_{i}
-\frac{x_{i}x_{j}}{x_{k}x_{k}}\pa_{i}\pa_{j}$ in Cartesian coordinates~\cite{hong00}.}

Next, we consider an eigenvalue equation of the form
\beq
\tilde{H}_{0}\psi(\theta,\phi)=E\psi(\theta,\phi).
\label{eigeneqn}
\eeq
Firstly, for a given angle $\theta$, we can have a reduced Sch\"rodinger equation
\beq
-\frac{\hbar^{2}}{2m(b+a\sin\theta)^{2}}\frac{d^{2}\psi}{d\phi^{2}}=E\psi
\eeq
to yield the eigenfunctions
\beq
\psi_{l}(\phi)=\frac{e^{il\phi}}{(2\pi)^{1/2}},
\eeq
with $l=0,\pm 1, \pm 2,...$ and the energy spectrum of the particle zero modes
\beq
E_{l}(\theta)=\frac{\hbar^{2}l^{2}}{2m(b+a\sin\theta)^{2}}.
\eeq
Note that in the limit of $b\gg a$, we can obtain the $\theta$-independent form
\beq
E_{l}=\frac{\hbar^{2}l^{2}}{2I_{b}},
\label{spectl}
\eeq
which describes the particle motion, with quantum number $l$, rotating
on an axially circular orbit of radius $b$ along the $\phi$-direction.  Here,
the moment of inertia of the particle is given by
\beq
I_{b}=mb^{2},
\label{ib}
\eeq
and $l$ is the angular momentum quantum number of the corresponding operator $J_{\phi}$.
Note that the quantum operator $J_{\phi}$ is defined on the two-dimensional
$x_{1}$-$x_{2}$ plane to yield the quantum number $l^{2}$ of $\vec{J}_{\phi}^{2}$.
Moreover, the angular momentum operator $\vec{J}_{\phi}^{2}$ produces the quantum number,
instead of $l(l+1)$, $l^{2}$ which is a characteristic of two-dimensional rigid
rotator~\cite{hongann}.

Secondly, for a given angle $\phi$, we can find a Sch\"rodinger equation of the form
\beq
-\frac{\hbar^{2}}{2ma^{2}(b+a\sin\theta)}\frac{d}{d\theta}\left(
(b+a\sin\theta)\frac{d\psi}{d\theta}\right)=E\psi.
\label{schphi1}
\eeq

Setting
\beq
\psi_{n}(\theta)=e^{in\theta}\Theta(\theta),
\label{psi}
\eeq
we can decompose the second-order Sch\"rodinger equation (\ref{schphi1}) into two
ordinary differential equations
\beq
\frac{d^{2}\Theta}{d\theta^{2}}+\frac{a\cos\theta}{b+a\sin\theta}
\frac{d\Theta}{d\theta}+\left(\frac{2ma^{2}E}{\hbar^{2}}-n^{2}\right)\Theta=0,
\label{ode01}
\eeq
and
\beq
\frac{d\Theta}{d\theta}+\frac{a\cos\theta}{2(b+a\sin\theta)}\Theta=0,
\label{ode02}
\eeq
from which we can obtain the eigenfunctions with $n=0,\pm 1, \pm 2,...$
\beq
\psi_{n}(\theta)=\frac{(b^{2}-a^{2})^{1/4}}{(2\pi)^{1/2}}\frac{e^{in\theta}}{(b+a\sin\theta)^{1/2}},
\eeq
and the energy spectrum of the particle zero modes
\beq
E_{n}(\theta)=\frac{\hbar^{2}}{2m}\left[\frac{n^{2}}{a^{2}}-\frac{\cos^{2}\theta}{4(b+a\sin\theta)^{2}}
-\frac{\sin\theta}{2a(b+a\sin\theta)}\right].
\label{spect2}
\eeq
In the limit of $b\gg a$, we can obtain the eigenfunctions of the particle zero modes
\beq
\psi_{n}(\theta)=\frac{e^{in\theta}}{(2\pi)^{1/2}},
\eeq
and the corresponding energy spectrum
\beq
E_{n}=\frac{\hbar^{2}n^{2}}{2I_{a}},
\label{spect22}
\eeq
where the moment of inertia $I_{a}$ of the particle is given by
\beq
I_{a}=ma^{2},
\label{ia}
\eeq
and $n$ is the angular momentum quantum number of the corresponding operator $J_{\theta}$.
Note that, on the two-dimensional cross sectional constant-$\phi$ plane, the quantum operator
$J_{\theta}$ is well defined to produce the quantum number $n^{2}$ of $\vec{J}_{\theta}^{2}$,
as mentioned in (\ref{spectl}).  In fact, the energy spectrum in the limit of $b\gg a$ denotes the particle motion,
with angular momentum quantum number $n$, rotating on a circular orbit of radius $a$
along the $\theta$-direction.

Thirdly, for a general case of (\ref{eigeneqn}), we again set
\beq
\psi_{nl}(\theta,\phi)=e^{i(n\theta+l\phi)}\Theta(\theta),
\eeq
to yield the first-order differential equation (\ref{ode02}) and a second-order
differential equation
\beq
\frac{d^{2}\Theta}{d\theta^{2}}+\frac{a\cos\theta}{b+a\sin\theta}
\frac{d\Theta}{d\theta}+\left[\frac{2ma^{2}E}{\hbar^{2}}
-\frac{a^{2}l^{2}}{(b+a\sin\theta)^{2}}-n^{2}\right]\Theta=0.
\label{secdiff}
\eeq
After some algebra, we can find the eigenfunctions with the quantum
numbers $n=0,\pm 1, \pm 2,...$ and $l=0,\pm 1, \pm 2,...$
\beq
\psi_{nl}(\theta,\phi)=\frac{(b^{2}-a^{2})^{1/4}}{2\pi}
\frac{e^{i(n\theta+l\phi)}}{(b+a\sin\theta)^{1/2}},
\label{psig}
\eeq
and the energy spectrum of the particle zero modes
\beq
E_{nl}(\theta)=\frac{\hbar^{2}}{2m}\left[\frac{n^{2}}{a^{2}}
+\frac{4l^{2}-\cos^{2}\theta}{4(b+a\sin\theta)^{2}}
-\frac{\sin\theta}{2a(b+a\sin\theta)}\right].
\label{spect3}
\eeq
Note that the energy spectrum (\ref{spect3}) is the most general solution to the Schr\"odinger
equation (\ref{eigeneqn}) for the quantum Hamiltonian operator (\ref{toham}) and
the corresponding eigenfunctions (\ref{psig}) satisfy the following orthogonality condition
\beq
\int_{0}^{2\pi}{\rm d}\theta\int_{0}^{2\pi}{\rm d}\phi~
\psi_{nl}(\theta,\phi)\psi_{n^{\prime}l^{\prime}}(\theta,\phi)
=\delta_{nn^{\prime}}\delta_{ll^{\prime}}.
\eeq
In order to investigate the particle dynamics associated with
the energy spectrum structure in terms of the toric geometrical
parameters $a$ and $b$, we consider the simple case of $b\gg a$
to arrive at the eigenfunctions
\beq
\psi_{nl}(\theta,\phi)=\frac{e^{i(n\theta+l\phi)}}{2\pi},
\eeq
and the corresponding $\theta$-independent energy spectrum of the particle zero modes
\beq
E_{nl}=\frac{\hbar^{2}}{2}\left(\frac{n^{2}}{I_{a}}+\frac{l^{2}}{I_{b}}\right),
\label{enl}
\eeq
where $I_{a}$ and $I_{b}$ are moments of inertia defined in (\ref{ia}) and (\ref{ib}),
respectively.  Note that in the case of $(n,l)=(0,l)$ the energy spectrum (\ref{enl})
is reduced to that of (\ref{spectl}), while in the case of $(n,l)=(n,0)$ to that of
(\ref{spect22}).  In the case of the angular momentum quantum numbers $(n,l)$ associated
with the operators $J_{\theta}$ and $J_{\phi}$ discussed above, the energy
spectrum (\ref{enl}) describes the particle motion, rotating on a helix orbit of radius
$a$ along the $\theta$-direction and pointing toward the $\phi$-direction of the torus
of axial radius $b$.

\section{Symplectic structures}
\setcounter{equation}{0}
\renewcommand{\theequation}{\arabic{section}.\arabic{equation}}

In this section, we show that the Dirac brackets obtained in the previous section are
in full agreement with those in the symplectic approach~\cite{faddeev88} to our second-class
system.  We start with considering the symplectic analogue of the conventional Dirac approach.
The master Lagrangian given by $L_{0}$ in (\ref{lag}) is of the form
\beq
L^{(0)}=a_{\alpha}^{(0)}\dot{\xi}^{(0)\alpha}-V^{(0)}
\eeq
where
\beq
\xi^{(0)\alpha}=(\theta,p_{\theta},\phi,p_{\phi},p_{r}),~~~
a_{\alpha}^{(0)}=(p_{\theta},0,p_{\phi},0,0),
\label{xia}
\eeq
and $V^{(0)}$ is given by $H_{0}$ in (\ref{hc}).  The Euler-Lagrange equations then read
\beq
f_{\alpha\beta}^{(0)}\dot{\xi}^{(0)\beta}=K_{\alpha}^{(0)},
\label{fab0}
\eeq
where $f_{\alpha\beta}^{(0)}$ is the (pre-)symplectic form
\beq
f_{\alpha\beta}^{(0)}=\frac{\pa a_{\beta}^{(0)}}{\pa \xi^{(0)\alpha}}-
\frac{\pa a_{\alpha}^{(0)}}{\pa \xi^{(0)\beta}},
\eeq
and $K_{\alpha}^{(0)}$ is given by
\beq
K_{\alpha}^{(0)}=\frac{\pa V^{(0)}}{\pa \xi^{(0)\alpha}}
\eeq
Explicitly we obtain
\beq
f^{(0)}=\left(
\begin{array}{ccccc}
0 &-1 &0 &0 &0\\
1 &0  &0 &0 &0\\
0 &0  &0 &-1 &0\\
0 &0  &1 &0 &0\\
0 &0  &0 &0 &0\\
\end{array}
\right),~~~
K^{(0)}=\left(
\begin{array}{c}
-\frac{p_{\phi}^{2}r\cos\theta}{m(b+r\sin\theta)^{3}} \\

\frac{p_{\theta}}{mr^{2}} \\
0 \\
\frac{p_{\phi}}{m(b+r\sin\theta)^{2}}\\
\frac{p_{r}}{m}\\
\end{array}
\right).
\label{matrix0}
\eeq
It is evident that since ${\rm det}f^{(0)}=0$, the matrix $f^{(0)}$ is not
invertible.  In fact the rank of this matrix is $4$, so that there exists
infinity of zero-generation (left) zero mode $\nu_{\alpha}^{(0)}$ as follows
\beq
\nu^{(0)T}=(0,0,0,0,1),
\label{nu0}
\eeq
where the superscript $T$ stands for transpose.  Correspondingly, we have an infinity of
zero-generation constraint
\beq
\nu_{\alpha}^{(0)T}\frac{\pa V^{(0)}}{\pa \xi^{(0)\alpha}}=0
\eeq
which leads to the constraint $\Omega_{2}=p_{r}$ in (\ref{const22}).  Our new set of
first-generation dynamical variables are then given by
\beq
\xi^{(1)\alpha}=(\theta,p_{\theta},\phi,p_{\phi},p_{r},\rho),
\label{xia1}
\eeq
and the first-generation Lagrangian takes the form
\beq
L^{(1)}=a_{\alpha}^{(1)}\dot{\xi}^{(1)\alpha}-V^{(1)}
\eeq
where
\beq
a_{\alpha}^{(1)}=(p_{\theta},0,p_{\phi},0,0,\Omega_{2}),
\label{aalpha1}
\eeq
and
\beq
V^{(1)}=\frac{p_{\theta}^{2}}{2mr^{2}}+\frac{p_{\phi}^{2}}{2m(b+r\sin\theta)^{2}}.
\label{hc1}
\eeq
The Euler-Lagrange equations now takes the form
\beq
f_{\alpha\beta}^{(1)}\dot{\xi}^{(1)\beta}=K_{\alpha}^{(1)},
\label{fab1}
\eeq
where the first-generation symplectic form $f_{\alpha\beta}^{(1)}$ is given by
\beq
f_{\alpha\beta}^{(1)}=\frac{\pa a_{\beta}^{(1)}}{\pa \xi^{(1)\alpha}}-
\frac{\pa a_{\alpha}^{(1)}}{\pa \xi^{(1)\beta}},
\eeq
and $K_{\alpha}^{(1)}$ is given by
\beq
K_{\alpha}^{(1)}=\frac{\pa V^{(1)}}{\pa \xi^{(1)\alpha}}.
\eeq

We then obtain explicitly
\beq
f^{(1)}=\left(
\begin{array}{cccccc}
0 &-1 &0 &0 &0 &0\\
1 &0    &0 &0 &0 &0\\
0 &0    &0 &-1 &0 &0\\
0 &0    &1 &0 &0 &0\\
0 &0    &0 &0 &0 &1\\
0 &0    &0 &0 &-1 &0\\
\end{array}
\right),~~~
K^{(1)}=\left(
\begin{array}{c}
-\frac{p_{\phi}^{2}r\cos\theta}{m(b+r\sin\theta)^{3}} \\
\frac{p_{\theta}}{mr^{2}} \\
0 \\
\frac{p_{\phi}}{m(b+r\sin\theta)^{2}}\\
0\\
0\\
\end{array}
\right).
\label{matrix1}
\eeq
Moreover, the inverse $f^{(1)}_{-1}$ of the matrix $f^{(1)}$ is given by
\beq
f^{(1)}_{-1}=\left(
\begin{array}{cccccc}
0 &1 &0 &0 &0 &0\\
-1 &0    &0 &0 &0 &0\\
0 &0    &0 &1 &0 &0\\
0 &0    &-1 &0 &0 &0\\
0 &0    &0 &0 &0 &-1\\
0 &0    &0 &0 &1 &0\\
\end{array}
\right).
\label{matrix2}
\eeq

Now, let $F$ and $G$ be functions of the dynamical variables $\xi^{(1)\alpha}$.
We can then define generalized symplectic structures as
\beq
\{F,G\}^{*}=\frac{\pa F}{\pa\xi^{(1)\alpha}}f_{-1}^{(1)\alpha\beta}
\frac{\pa G}{\pa\xi^{(1)\beta}}.
\eeq
In particular, we have the following  symplectic structures
\beq
\{\xi^{(1)\alpha},\xi^{(1)\beta}\}^{*}=f_{-1}^{(1)\alpha\beta},
\eeq
to yield
\beq
\{\theta,p_{\theta}\}^{*}=1,~~~\{\phi,p_{\phi}\}^{*}=1,~~~\{\rho,p_{r}\}^{*}=0.
\label{sbracket}
\eeq
Note that, with the identification $\rho=r$, the symplectic brackets in
(\ref{sbracket}) reproduce the Dirac brackets in (\ref{dbracket}).  Moreover, together with
the matrices (\ref{matrix1}), the Euler-Lagrange equations (\ref{fab1}) reconstruct
the canonical momenta (\ref{cojm}), to yield
\beq
\dot{p}_{\theta}=\frac{p_{\phi}^{2}r\cos\theta}{m(b+r\sin\theta)^{3}}
=mr(b+r\sin\theta)\cos\theta\dot{\phi}^{2},~~~
\dot{p}_{\phi}=0
\eeq
attainable also from the variations of the Lagrangian (\ref{lag}) with
respect to the coordinates $\theta$ and $\phi$, and the consistency conditions of
the constraints associated with the coordinate $r=\rho$
\beq
\dot{\Omega}_{1}=0,~~~\dot{\Omega}_{2}=0.
\eeq

\section{Conclusion}

In conclusion, we have constructed a first-class Hamiltonian by
exploiting the St\"uckelberg coordinates associated with the
geometrical constraints imposed on the torus.  Subsequently, in the BFV scheme
we have obtained the BRST-invariant gauge fixed Lagrangian including
the ghosts and anti-ghosts, and the BRST transformation rules under which the
effective Lagrangian is invariant.  We have also discussed the particle dynamics by
constructing the spectrum  of the free particle system on the torus.   The
symplectic structures involved in the system on the torus were also
investigated to yield the consistency with the Dirac analysis for the second-class
constraints. It would be interesting to study a field theoretic extension of
this work with toric geometric constraints.

\acknowledgments
The author would like to acknowledge financial support in part
from the Korea Science and Engineering Foundation grant R01-2000-00015.

\end{document}